\documentclass[12pt]{article}
\usepackage[utf8]{inputenc}
\usepackage{graphicx}
\usepackage{longtable}
\usepackage{hyperref}
\usepackage{amsmath}
\usepackage{geometry}
\usepackage{array}
\usepackage{parskip}
\usepackage{url}
\usepackage{enumitem}
\usepackage{multirow}
\usepackage{adjustbox}
\usepackage{multicol}

\begin{document}
\title{\textbf{Harnessing the Potential of Gen-AI Coding Assistants in Public Sector Software Development}}

\author{
    Kevin KB Ng, Liyana Fauzi, Leon Leow, Jaren Ng \\
    {\small \textit{Emails: kevin\_kb\_ng@tech.gov.sg, liyana\_muhammad\_fauzi@tech.gov.sg,}} \\
    {\small \textit{leon\_leow@tech.gov.sg}} \\\\
    \text{Government Technology Agency (GovTech)} \\
    Singapore
}

\date{April 2024}
\maketitle

\begin{abstract}
   The study on GitHub Copilot by GovTech Singapore’s Engineering Productivity Programme (EPP) reveals significant potential for AI Code Assistant tools to boost developer productivity and improve application quality in the public sector. Highlighting the substantial benefits for the public sector, the study observed an increased productivity (coding / tasks speed increased by 21-28\%), which translates into accelerated development, and quicker go-to-market, with a notable consensus (95\%) that the tool increases in developer satisfaction. Particularly, junior developers experienced considerable efficiency gains and reduced coding times, illustrating Copilot's capability to enhance job satisfaction by easing routine tasks. This advancement allows for a sharper focus on complex projects, faster learning, and improved code quality. Recognising the strategic importance of these tools, the study recommends the development of an AI Framework to maximise such benefits while cautioning against potential over-reliance without solid foundational programming skills. It also advises public sector developers to classify their code as "Open" to use Gen-AI Coding Assistant tools on the Cloud like GitHub Copilot and to consider self-hosted tools like Codeium or Code Llama for confidential code to leverage technology efficiently within the public sector framework. With up to 8,000 developers, comprising both public officers and vendors developing applications for the public sector and its customers, there is significant potential to enhance productivity.
\end{abstract}

\section{Introduction}
\subsection{Literature Review}
Artificial Intelligence (AI) coding assistants have increasingly become vital tools for improving software development productivity. General-purpose AI assistants, such as GitHub Copilot, have shown promise in automating routine tasks like code completion and documentation, as noted by Gold, Smith, and Davis (2023)\cite{Gold2023}. These tools help developers reduce time spent on repetitive tasks, allowing them to focus on more complex aspects of coding. However, general-purpose assistants face limitations when it comes to handling domain-specific requirements, particularly in highly regulated sectors like digital public sector, where security and compliance are critical (Lee, Wang, \& Davis, 2021\cite{Lee2021}; Mansour, Wu, \& Nguyen, 2022\cite{Mansour2022}). Other studies have noted that productivity gains come at the expense of reduced code quality (S.Imai, 2022\cite{Imai2022}) and potentially less secure code (Perry et al. 2023\cite{Perry2023}; Elgedawy et al. 2024\cite{Elgedawy2024}). Understanding and managing the drawbacks and limitations of coding assistants may be key to harnessing their potential.

Schmidt et al. (2024)\cite{Schmidt2024} provided a broad overview of the adoption of AI-based coding assistants in practice. Their study highlights varying degrees of adoption across different stages of the software development lifecycle, with developers more comfortable using AI assistants for tasks such as writing tests and generating documentation while hesitating to use them for more complex and sensitive tasks like refactoring or bug triaging. The study also pointed to a general lack of trust in AI-generated code, particularly around security, which limits widespread adoption. This is echoed by several large-scale surveys of 400+ developers (Liang et al. 2024\cite{Liang2024}; Sergeyuk et al. 2024\cite{Sergeyuk2024})  that noted motivation for using coding assistants was primarily aimed at reducing time spent on repetitive tasks from speeding up keystrokes and recalling syntax to writing unit tests but not on solutioning – suggesting that the current state of coding assistants in 2024 may be better positioned as a baseline tool to accelerate code synthesis as opposed to a solution engine for developers to delegate complex contextual, security and logical challenges in software development.

To address these challenges, researchers like Pinto et al. (2023)\cite{Pinto2023} and Wu et al. (2023)\cite{Wu2023} have explored contextualised AI coding assistants that incorporate domain-specific knowledge and internal APIs. Pinto et al.'s "StackSpot AI" demonstrated the potential for contextualised assistants to improve the accuracy and efficiency of developers working with proprietary systems, such as those found in digital public sector environments. These tools can generate accurate code and documentation aligned with internal requirements, helping developers manage large and complex codebases more effectively. However, as Wu et al. (2023)\cite{Wu2023} noted, challenges remain in ensuring these tools integrate well with existing workflows, provide reliable outputs, and address security concerns. Tan, et al. (2024)\cite{Tan2024} suggest that supplementing contextual information in annotations and optimising interactions with coding assistants may be key to maximise its effectiveness and limits. With future developments in coding assistants, contextual awareness is expected to increase. Addressing the immediate challenges of security, trust, context and maximising effectiveness in the public sector context presents an opportunity to harness productivity benefits ahead of technological advancements.

In the context of digital public sector, AI coding assistants offer the potential to significantly enhance software development by automating routine tasks and providing developers with more time to focus on innovation and problem-solving. However, successful deployment hinges on overcoming challenges related to security, trust and maximising AI tools’ ability to understand specific requirements of public sector agencies. This paper will explore how to effectively harness AI-driven tools to support the unique demands of software development in the digital public sector.

\subsection{Developer Productivity}

\textbf{1.1.1} 
EPP\footnote{EPP is a 5-year whole-of-public sector effort, led by Smart Nation Singapore and GovTech, to transform the use of infrastructure for system modernisation, improved individual developer productivity and increased system resiliency.}is committed to boosting the productivity of developers – comprising both public officers and external vendors - working for the public sector. This strategic emphasis is designed to free up developers' time for more impactful work, thereby supporting public sector agencies in their mission to provide citizens, businesses, and public servants with secure, high-quality applications. The adoption of AI technology across various fields is seen as crucial and is expected to become commonplace. AI's ability to enhance efficiency, reduce mistakes, and possibly automate monotonous tasks at different points in the software development process is particularly highlighted.

\textbf{1.1.2} 
To assess the effectiveness of coding assistant tools in this context, the Government Digital Products division in GovTech Singapore conducted a pilot study under the SHIP-HATS\footnote{SHIP (Secure Hybrid Integration Pipeline)-HATS (Hive Agile Testing Solutions) is a Continuous Integration/Continuous Delivery CI/CD component within SG Government Tech Stack (SGTS) with security and governance guardrails that enables developers to plan, build, test, and deploy code to production.} initiative, a part of the Singapore public sector’s DevSecOps framework and technology stack. The study focused on GitHub's Copilot, an AI-enhanced coding assistant developed in partnership between GitHub and OpenAI, known for its ability to offer developers real-time code suggestions and completions. An examination of GitLab Duo was also conducted to assess its utility.

\subsection{Potential of AI Coding Assistants in the public sector context}

\textbf{1.2.1}
The value of coding assistant tools is especially pronounced in the context of public sector-led projects, such as those that utilise SHIP-HATS for development, testing, and deployment. These projects are often constrained by tight deadlines, limited budgets, and stringent requirements for security and compliance. Coding assistant tools, like GitHub Copilot, can play a pivotal role in addressing these challenges. They have the potential to expedite the development process, enhance the quality of code, and free developers to concentrate on crafting secure, efficient, and user-centric public services. Integrating such AI-driven tools into public sector software development signifies a shift towards increased productivity and marked improvement in the delivery of public services.

\textbf{1.2.2}
Coding assistant tools, such as GitHub Copilot, play a crucial role in enhancing developer productivity for several compelling reasons. These tools leverage advanced technologies, including AI and Machine Learning (ML), to automate routine tasks, suggest code snippets, and even generate entire functions based on the context provided by the developer. These functionalities present the potential to alleviate the challenges in several ways:
\begin{enumerate}[label=\alph*)]
\item \textbf{Time Savings:} Coding assistants significantly reduce the time developers spend on repetitive tasks by automating routine coding tasks and suggesting code snippets. This allows developers to focus on more complex, high-value aspects of their projects, accelerating the development cycle.

\item \textbf{Reduced Errors:} They can help minimise human errors by suggesting tested and verified code snippets. This leads to cleaner, more reliable codebases and reduces the time spent on debugging and error correction, enhancing overall code quality.

\item \textbf{Learning and Development:} Coding assistants can serve as an educational tool for less experienced developers, offering suggestions and examples that can help them learn new programming languages or frameworks more quickly. This accelerates the onboarding of new technologies and enhances the team’s overall technical skill level within the team.

\item \textbf{Enhanced Collaboration:} Coding assistants can improve the consistency and readability of code by standardising code snippets and ensuring that suggested code follows best practices. This makes it easier for teams to collaborate as code written by different developers becomes more uniform and understandable.

\item \textbf{Increased Innovation:} With routine tasks handled by AI, developers can allocate more time to innovation and exploring new technologies or approaches. This can lead to the development of new features, products, and services that might not have been possible under traditional development workflows.

\item \textbf{Scalability:} Coding assistants can easily scale with the team, providing consistent support regardless of team size. This scalability ensures that the benefits of coding assistants extend across the organisation, from small teams to large enterprises.
\end{enumerate}

\textbf{1.2.3}
Industry reports suggest an impressive 30-100\%\footnote{Deniz, B. K., Gnanasambandam, C., Harrysson, M., Hussin, A., \& Srivastava, S. (2023, June 27). Unleashing developer productivity with Generative AI. McKinsey \& Company. https://www.mckinsey.com/capabilities/mckinsey-digital/our-insights/unleashing-developer-productivity-with-generative-ai} increase in coding speed. Beyond speed, these tools lessen the cognitive burden of routine tasks on developers, allowing for greater focus on creative and problem-solving activities. This shift not only boosts job satisfaction but also encourages innovation within projects, demonstrating the substantial and multifaceted advantages of coding assistant tools in the public sector’s software development landscape.

\section{Pilot Study Methodology}
\subsection{Focusing on the “Development” Stage}
AI can benefit all phases of the software development lifecycle (see Figure 1). For example, during the requirements stage, plugins in tools like Jira use generative AI to automate the creation of user stories by providing AI-driven assistance in defining project scope, setting achievable goals, and generating user stories, traditionally a manual process. Similarly, in the testing stage, tools like Fortify Audit Assistant identify relevant exploitable vulnerabilities in your organisation in new static scan results. This is achieved through scan analytics machine-learning classifiers trained using anonymous metadata from scan results previously audited by software security experts. 

\begin{figure}[htbp]
    \centering
    \includegraphics[width=0.8\textwidth]{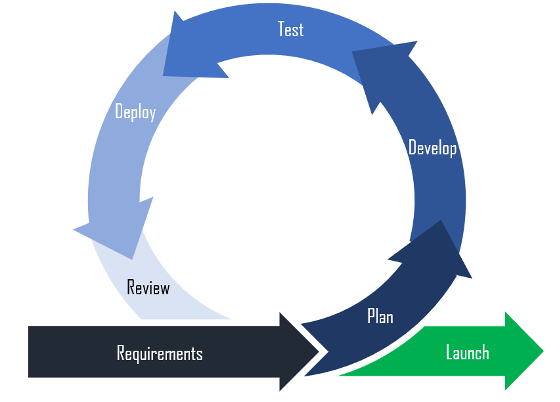}
    \caption{Agile Software Development Lifecycle}
    \label{fig:SDLC}
\end{figure}

While AI can help various phases of the software development lifecycle, the development phase stands out due to its complexity, iterative nature, and the need for creativity. Code assistant tools can automate complex tasks, provide insights for iterative improvements, enhance creativity, and help developers stay updated with the latest trends, including advancements in programming languages, frameworks, tools, best practices, and security protocols.

\subsection{Study Scope}
\normalsize
The study utilised GitHub Copilot for Business\footnote{GitHub Copilot Business offers advanced features over the individual plan, including enhanced AI model for better code suggestions, AI-based security vulnerability filtering, VPN proxy support for flexible working environments, and administrative features for managing access and policies within an organisation} over four months from Oct ’23 to Jan ’24. GitHub Copilot is platform-agnostic and can be used with any Source Code Management and CI/CD Platform. For this pilot study, participants could use GitHub Copilot with SHIP-HATS’ GitLab. To use GitHub Copilot, participants installed it directly into their Integrated Development Environments (IDEs) of choice. All participants were required to use GitHub Copilot in their day-to-day work. 

\subsection{Participant Selection}
Invitations to the pilot were sent to current SHIP-HATS users actively involved in developing apps for the public sector. A total of 70 developers signed up for the pilot. While there was a range of participants, varying in years of experience, most were Junior and Mid-level Software Engineers (SWEs) who make up about three-quarters of GovTech’s SWE population.

\subsection{Data Collection and Analysis}
\textbf{2.5.1}
A combination of \textbf{telemetry data} (e.g., prompt acceptance rates) and \textbf{survey data} was used to assess Copilot’s impact on developers’ productivity and to gather insights into the overall developer experience.

\textbf{2.5.2}
\textbf{Telemetry data} on prompts and accepted lines of code from GitHub Copilot was collected. Prompts refer to user inputs, such as comments, partial code snippets, or a combination of both, that guide the AI in generating code suggestions. The quality and clarity of prompts greatly influence the relevance and usefulness of GitHub Copilot's outputs. “Accepted Lines-of-Code" represents the code suggestions from GitHub Copilot that users choose to integrate into their projects, reflecting the value and alignment of the AI's suggestions with the users’ needs.

\textbf{2.5.3}
\textbf{Survey data} was collected using a questionnaire based on the SPACE framework (see Annex B), drawing on GitHub’s experience with similar surveys (see Annex C). Industry benchmarks were also obtained to compare nuanced similarities or differences in the public sector’s developers’ experience against that of other developers in general. User interviews were also planned to explore survey findings further and gain a deeper understanding of the participants’ perspectives.

\section{Results}
\subsection{Telemetry Data Findings}

\textbf{3.1.1}
The telemetry data findings were based on the 70 pilot participants. The study (see Table 1) saw an \textbf{Acceptance Rate of 22\%}, with Typescript and Terraform receiving the highest acceptance rate, close to 30\%, while Java has less than 5\% acceptance rate. A higher acceptance indicates that users generally find the suggestions from GitHub Copilot to be accurate, useful, and reliable. It suggests that the tool effectively understands the context of the coding task and offers relevant solutions. In contrast, a lower acceptance rate suggests that users might find the suggestions less relevant, accurate, or useful. This could point to areas where the tool needs improvement in understanding user intent or handling complex coding tasks. 

\textbf{3.1.2}
As the number of users per programming language is not controlled in this pilot 	study, the pilot is inconclusive if any programming language will benefit more 	from GitHub Copilot. This is because the AI model behind GitHub Copilot has been trained on a vast amount of code written in these languages, allowing it to provide more accurate and relevant suggestions.

\textbf{3.1.3}
On the choice of programming languages and AI, in a structure made up of multiple organisations such as the Singapore public sector, a standardised tech stack across all public sector developers could yield significant benefits. This \textbf{uniformity could facilitate valuable cross-learning}, which is essential for developing ML models that power AI tools.

\begin{table}[h!]
\centering
\begin{tabular}{|>{\raggedright\arraybackslash}p{0.2\linewidth}|>{\centering\arraybackslash}p{0.2\linewidth}|>{\centering\arraybackslash}p{0.2\linewidth}|>{\centering\arraybackslash}p{0.2\linewidth}|}
\hline
\textbf{Language} & \textbf{Cumulative Number of Accepted Prompts} & \textbf{Acceptance Rate} & \textbf{Cumulative Number of Accepted Lines of Code} \\ \hline
Terraform         & 395                                         & 30.38\%& 1,223                                              \\ \hline
Typescript        & 4,021                                       & 27.82\%                                & 6,657                                              \\ \hline
Javascript        & 311                                         & 23.33\%                                & 474                                                \\ \hline
Hcl-terraform     & 59                                          & 23.32\%                                & 154                                                \\ \hline
Go                & 136                                         & 23.21\%                                & 168                                                \\ \hline
Kotlin            & 776                                         & 22.46\%                                & 1,805                                              \\ \hline
Python            & 771                                         & 20.56\%                                & 771                                                \\ \hline
Typescriptreact   & 767                                         & 18.63\%                                & 1,238                                              \\ \hline
JSON              & 102                                         & 16.59\%                                & 184                                                \\ \hline
Terragrunt        & 166                                         & 16.40\%                                & 316                                                \\ \hline
Ruby              & 234                                         & 16.05\%                                & 295                                                \\ \hline
Yaml              & 197                                         & 13.89\%                                & 297                                                \\ \hline
Shellscript       & 43                                          & 13.19\%                                & 55                                                 \\ \hline
Java              & 49                                          & 4.97\%                                 & 118                                                \\ \hline
\end{tabular}
\caption{Programming Language Acceptance and Code Statistics}
\end{table}

\textbf{3.1.4}
For \textbf{Accepted Lines of Code}, Typescript, Kotlin, and Typescriptreact have the most accepted lines of code, generally indicating a greater usefulness	acceptance of the code suggestions provided by GitHub Copilot. This could 	mean 	that the developers in the public sector are finding GitHub Copilot's suggestions 	relevant and efficient, potentially speeding up the coding process by reducing the 	 time and effort required to write code from scratch. It could also imply that the code generated by GitHub Copilot is of high quality, meaning it is reliable, efficient, and perhaps even following best practices, leading developers to accept the suggestions more often. Lastly, a higher number of accepted lines indicates that GitHub Copilot effectively understands the context of the development 	tasks and generates more accurate suggestions that align with the developers' needs.

\subsection{Survey Findings – SPACE Framework}
40 respondents, predominantly from GovTech, responded to the survey. See Annex D for detailed survey responses. 

\subsubsection{Satisfaction \& Wellbeing}

\begin{enumerate}[label=\alph*)]
\item \textbf{Improved Developer Satisfaction:}
The majority of respondents \textbf{(95\%) agreed that they could focus on more satisfying work} with coding assistants such as GitHub Copilot (which is higher than the industry response of 60 – 75\%) and that they felt more fulfilled at their jobs. In comparison, 60\% agreed that coding assistants are essential tools for improving developer productivity.  
The key reasons for improved levels of satisfaction and fulfilment from using GitHub Copilot are the reduction of repetitive tasks (e.g., writing unit tests) and the auto-completion of code.
A further 90\% of respondents agreed that the department should provide coding assistant tools, suggesting that such tools may eventually become a norm in a developer’s repertoire of tools.

\item \textbf{Increased Learning Acceleration:}
On average, three-quarters of respondents agreed that GitHub Copilot helped them learn new languages (see Figure 2). This suggests that developers are not only coding faster but are also scaling their skills more rapidly with the aid of AI-driven insights and examples. In interviews, some senior engineers gave feedback that if the developer does not have strong fundamentals or is new to the craft, learning through coding assistants may compound mistakes or lead to bad habits due to the occasional hallucinations of the machine learning model.
\end{enumerate}

\begin{figure}
    \centering
    \includegraphics[width=0.75\linewidth]{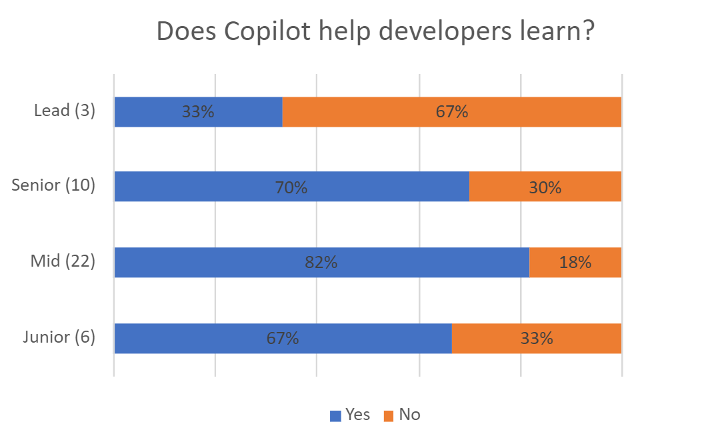}
    \caption{Learning Acceleration with Github Copilot}
    \label{Learning_Acceleration}
\end{figure}

\subsubsection{Performance}
The study highlights an enhancement in code quality, as developers reported fewer coding mistakes and higher code standards. The majority (\(>70\%\)) observed a reduction in coding mistakes and noted an improvement in code quality underscoring the impact of AI tools in producing cleaner, more reliable code. This improvement is likely attributed to the real-time feedback and suggestions provided by the AI, guiding developers towards best practices and away from potential errors. 

While the majority acknowledged the reduction in coding mistakes and the improved code quality with GitHub Copilot, in user interviews, there was still a strong sentiment that the accuracy of GitHub Copilot can be further improved.

\subsubsection{Activity}
Respondents provided feedback indicating that the \textbf{average time spent on coding was reduced by 22\%, equating to a 28\% increase in coding speed.} Furthermore, 88\% of respondents expressed an improved ability to complete their tasks within the expected time, with most confirming that the tool aided in task completion. Coding time was notably reduced for Junior and Mid-level developers, who often encounter more repetitive coding work. Nevertheless, there is compelling evidence that GitHub Copilot/GitLab Duo helps reduce the time spent on coding activities beyond repetitive tasks. \textbf{This aligns with the industry's observed coding speed increase (30-100\%), although not to the same extent.} Additionally, in a study on "The Impact of AI on Developer Productivity: Evidence from GitHub Copilot"\footnote{For more information, see: Van Aken, D., Sadler, G., Xie, K., and Zhang, L. (2023). *Visions of Autonomous AI: Survey of Recent Advances and Challenges*. Retrieved from \url{https://arxiv.org/abs/2302.06590}}. by Cornell University, it was found that in a controlled experiment, a group using GitHub Copilot completed a task 55\% faster than the group that did not use GitHub Copilot, and 78\% of the Copilot group fully completed the task compared to 70\% in the non-GitHub Copilot group. The conclusion we can draw is that the use of AI, specifically GitHub Copilot, has been observed to significantly increase coding speed and productivity in the industry, as well as in the public sector.

\subsubsection{Communication and Collaboration}
Feedback from the respondents reveal that generative AI tools like GitHub Feedback from the respondents reveals that generative AI tools like GitHub Copilot are valued for their ability to assist in code documentation and enhance the understanding of code among developers. Despite these positive aspects, there is a divide in opinions regarding their efficacy in reducing the time spent reviewing merge requests, with over half of the respondents indicating no reduction in time spent on these reviews. Additionally, while GitHub Copilot has been credited with improving documentation, aiding in better comprehension of others' code during hand-offs, and fostering a more cohesive development environment, challenges remain. Developers pointed out the tool's limitations in grasping the context of code, merely reflecting its functionality based on syntax, which suggests a need for further refinement. Nevertheless, a significant 66\% of respondents acknowledged GitHub Copilot's contribution to better documentation in code, underscoring its potential to streamline collaboration and communication, albeit with room for improvement in understanding code context and enhancing collaborative development processes.

\subsubsection{Efficiency and Flow}
Respondents perceived\textbf{ a notable 24\% productivity increase} across the board when it comes to coding, with junior developers benefitting the most from this enhancement, as opposed to their senior or lead counterparts. This uptick in productivity is closely linked with accelerated coding speed. Respondents reported a decrease in mental effort for repetitive tasks and a reduction in the time needed to search for information. However, user interviews shed light on the tool's limitations in handling complex tasks that require a deep understanding of contextual details, like the intricacies of microservices or APIs. Despite the overall decrease in mental load for repetitive coding tasks reported by a high number of respondents and three-quarters experiencing less time spent on information lookup, over half of the participants noted no improvement in the efficiency of hand-offs. These mixed results highlight the complex relationship between tool use and productivity, emphasising the importance of further investigation into the tool's ability to meet the complex demands of software development.

\begin{table}[h!]
\centering

\begin{tabular}{|>{\raggedright\arraybackslash}p{0.2\linewidth}|>{\centering\arraybackslash}p{0.2\linewidth}|>{\centering\arraybackslash}p{0.2\linewidth}|>{\centering\arraybackslash}p{0.2\linewidth}|}
\hline
\textbf{Experience Level} & \textbf{Coding Time Reduction with GitHub Copilot} & \textbf{Task Completion Speed Increase with GitHub Copilot} & \textbf{Productivity Increase with GitHub Copilot} \\ \hline
Junior (6) – Starting out in the field & 33\% & 34\% & 44\% \\ \hline
Mid-Level (22) – Several years of experience & 24\% & 21\% & 24\% \\ \hline
Senior (10) – Many years of experience and expertise in specific domains & 15\% & 15\% & 17\% \\ \hline
Lead (3) – Senior and leading others & 13\% & 8\% & 12\% \\ \hline
Total Average & 22\% & 21\% & 24\% \\ \hline
\end{tabular}
\caption{Coding, Task and Productivity improvements from study}
\end{table}

\subsubsection{Security}
We reviewed the impact of using GitHub Copilot on code security, finding that many users did not observe significant enhancements despite the expected improvement in early vulnerability detection. This suggests that while AI tools are revolutionising development, their real-time capability to identify potential security threats is still developing.  It may clarify known vulnerabilities but not effectively detect them on the fly compared to SAST scanners. Developers may need to rely on AI components in security testing tools like Fortify-on-demand for more advanced vulnerability detection and remediation.

\section{Pilot Results Summary}

\subsection{Coding Speed \& Developer Satisfaction}	
4.1.	The pilot study highlighted improvements in satisfaction, workflow efficiency, and overall productivity among developers, with the observed average decrease of coding time by 22\% except among junior and mid-level developers, who observed a higher 33\% and 24\% decrease in coding time respectively. This is largely due to more repetitive coding tasks, such as writing unit tests encountered at more junior levels. Similarly, the productivity gains averaged a 24\% increase, with junior developers benefiting more significantly. It is, however, presumed that half of the productivity gains are attributed to coding time reduction and the other half to adjacent efforts such as using Copilot for solutioning, creating pull requests, generating code comments / documentation etc. A smaller independent trial done by another team in GovTech yielded similar positive coding reduction time as well at 30\%\footnote{In the Evaluation Report on GitHub Copilot Trial for PIC (Product Innovation Center), the team trialing GitHub Copilot observed a 30\% reduction in task completion time and 80\% found the tool able to help them to better focus on satisfying work.}.

\subsection{Concerns over Learning, Accuracy and Security}
Some concerns emerged regarding the learning curve associated with GitHub Copilot. Senior developers expressed apprehension that the tool's occasional inaccuracies, or "hallucinations," might encourage poor coding practices or the adoption of incorrect information among less experienced developers. Furthermore, the tool did not effectively address security vulnerabilities as had been claimed.

\subsection {Overall Developer Productivity}
With developers \textbf{reporting spending on average 50\% of their time doing development tasks in the IDE (Integrated Development Environment) where coding assistants have an impact on, the 24\% observed coding productivity translates roughly into a 12\% overall developer productivity gain} when using coding assistants. In summary, the study observed that the use of GitHub Copilot showcased meaningful improvements in public sector developer productivity by 12\% using coding assistants which can lead to a potential time savings of approximately 5 hours per week, or 30 man-days annually per developer - assuming that across developers of differing levels, each is spending approximately 4 hours per day doing development tasks in the IDE. 

In the long run, GitHub Copilot and similar coding assistant tools are likely to increase public sector developer productivity even more than what is observed in this study as tool familiarity improves and the large language models powering the coding assistants are enhanced further.

\section{Recommended Actions to enable effective use of Coding Assistants in Public Sector Context}
With the study concluding, to enable public sector developers to harness the benefits of AI coding assistants like GitHub Copilot in the short-term, the following will be considered:

\subsection{Maximising benefits of Code Assistants}
In the immediate term, before optimised and public sector tailored solutions are available, agencies public sectors can consider the following approach to effectively harness the benefits of coding assistants: 

\begin{enumerate}[label=\alph*)]
    \item Review the context of data security against coding assistant use cases and select the appropriate implementation of coding assistants (SaaS, self-hosted in an air-gapped environment such as Code Llama\footnote{Code Llama, created by Meta, is an open-source, Llama 2-based, coding LLM designed for generating and understanding code across various languages, available for both cloud and self-hosted implementations.}, etc.).
    
    \item Develop playbooks and guidelines around the usage of coding assistants that align with the daily work of developers to both ring-fence acceptable usage scenarios and take advantage of productivity benefits while mitigating pitfalls.
    
    \item Provide support avenues and channels for sharing information, developing practices, and reporting potential malpractices or leaks of classified data.
\end{enumerate}

\subsection{Optimize tooling approach and develop better usage practices}
The productivity gap with the broader industry might be due to limited tool exposure. Adopting best coding practices and AI coding assistants could align our productivity with industry standards. Investigating this further will help identify improvement opportunities. For better technology use across the public sector, it is worth exploring the viability of a standard tech stack. A standard tech stack allows for collective wisdom and sharing of insights but also ensures the creation of secure, quality software due to repeatable coding snippets and norms. Also, providing targeted AI tool training for public sector developers (e.g. prompt engineering) is crucial to maximise benefits and streamline public sector tech operations more effectively.

\subsection{Formulate policy framework to address security and compliance}
Further assessment of how these coding assistants can be enhanced to align with the public sector’s stringent security and compliance requirements is necessary to ensure the integrity and safety of public sector software solutions. The creation of comprehensive policies for the ethical and effective adoption of AI coding assistants, focusing on inclusiveness, accessibility, and compliance with national security and data protection regulations, would guide the public sector’s usage of gen-AI tools.

\raggedright
\section{Relationship with industry standards such as DORA and SPACE}
DORA (DevOps Research and Assessment) and SPACE are industry frameworks that are pivotal to measuring developer productivity, with SPACE targeting individual developers and DORA at the team level through metrics like Deployment Frequency, Lead Time for Changes, Time to Restore Service, and Change Failure Rate. These metrics offer insights into the speed and reliability of software delivery, key for refining DevOps practices. Through this pilot, the SPACE framework has offered crucial insights into our engineering productivity and work environment, supporting the decision to continue utilizing GitHub Copilot and other AI coding assistants. Future evaluations should leverage DORA metrics alongside SPACE findings, to comprehensively assess coding assistants’ effect in improving team development productivity as part of EPP. Adopting this multifaceted approach ensures that the implementation of GitHub Copilot and similar technologies significantly boosts productivity, efficiency, and overall software delivery outcomes, validating the decision to recommend GitHub Copilot through both qualitative and quantitative evidence.

\newpage
\section{Conclusion}
In conclusion, our pilot findings reveal that AI coding assistants like GitHub Copilot offer significant opportunities to enhance developer productivity in public sector projects, demonstrating a coding/task speed increase of 21-28\%. Although slightly below the industry's broader observations of 30-100\% due to challenges such as varying expertise levels and limited usage because of in-country data requirements, this improvement emphasises a high potential for substantial efficiency gains. The observed coding speed increase is projected to translate to an overall improvement in developer productivity by 12\%. Case in point for improved developer productivity: in a recent project, the integration of Copilot increased a small team’s development productivity by 50\%\footnote{ The Role-Informer-Tool Project team in GovTech’s Government Digital Products Division estimated a 50\% increase in productivity where the four-FTE team was able to complete the work required of 2 more FTEs by leveraging AI for tasks such as UX, code generation, refactoring, and testing.}, resulting in quicker service delivery to the public. To benefit from coding assistants, agencies should review if their code could be right-classified to utilise cloud-based coding assistants such as GitHub Copilot as a next step. If not, self-hosted options like Code Llama are a possible alternative. GovTech will also facilitate access to different coding assistants, potentially through centrally procured licenses for public sector developers. With around 8,000 developers engaged in public sector applications, the scale of potential productivity improvement is immense. As we move forward, we aim to refine the governance and practices of AI coding assistants to ensure their compliant and effective use. 

As the software industry continue to move towards a customer-centric orientation, fast deployment becomes essential - a goal that AI coding assistants are well-positioned to facilitate. By strategically adopting these tools, we not only enhance operational efficiency but also position our public sector at the forefront of technological innovation. 
We invite collaboration and feedback from all public sector agencies to maximise the benefits of AI coding assistants, ensuring our collective advancement towards streamlined development processes and improved service delivery in a public sector context.

\newpage
\appendix
\section{Annex A: References}

\section{Annex B: SPACE Framework}
The SPACE framework is an approach for measuring, understanding, and optimizing engineering productivity, specifically within the context of developer work. It is an acronym that stands for Satisfaction, Performance, Activity, Communication, and Efficiency. These five dimensions are essential for a comprehensive understanding of productivity in development teams.

1.	\textbf{Satisfaction and Well-being:} This dimension focuses on the happiness and contentment of developers, considering their mental and physical health as crucial components of productivity.

2.	\textbf{Performance:} This aspect evaluates how effectively developers complete tasks, deliver projects, and achieve their goals, emphasizing the quality and speed of their output.

3.	\textbf{Activity:} This measures the actual work done by developers, including coding, debugging, and other tasks directly related to software development.

4.	\textbf{Communication and Collaboration:} This dimension assesses how developers interact with each other, share knowledge, and work together on projects, recognizing the importance of teamwork in achieving objectives.

5.	\textbf{Efficiency and Flow:} This focuses on how smoothly developers can work, with minimal disruptions and maximum productivity, emphasizing the importance of creating environments that support concentrated work.

The SPACE framework was outlined by researchers from GitHub, Microsoft, and the University of Victoria. It encourages leaders to view productivity holistically, placing metrics in context with each other and linking them to team goals rather than individual effort. This framework is well-regarded in the industry for its comprehensive approach to understanding and optimizing productivity in software development teams.

\section{Appendix C: Survey Question Guidance from GitHub}

\subsection{Survey Questions}

1. What is your user type?
    \begin{enumerate}[label=\alph*)]
        \item Administrator
        \item Assigned User
    \end{enumerate}

2. Which development environment do you use?
    \begin{enumerate}[label=\alph*)]
        \item Visual Studio Code
        \item Visual Studio
        \item JetBrains
        \item Neovim
        \item Other (Please specify)
    \end{enumerate}

3. What programming languages do you usually use?
    \begin{enumerate}[label=\alph*)]
        \item Java
        \item JavaScript
        \item Python
        \item TypeScript
        \item Ruby
        \item Go
        \item C\#
        \item Rust
        \item HTML
        \item Other (Please specify)
    \end{enumerate}

4. What best describes your programming experience?
    \begin{enumerate}[label=\alph*)]
        \item Student / Intern learning to program
        \item 0 to 2 years of professional programming experience
        \item 3 to 5 years of professional programming experience
        \item 6 to 10 years of professional programming experience
        \item 11 to 15 years of professional programming experience
        \item More than 16 years of professional programming experience
    \end{enumerate}

5. In which of the following ways did you use GitHub Copilot?
    \begin{itemize}
        \item For everyday coding work in a familiar language
        \item For everyday coding work in an unfamiliar language
        \item To write particularly repetitive code (boilerplate) in a familiar language
        \item To write particularly repetitive code (boilerplate) in an unfamiliar language
        \item To write tests in a familiar language
        \item To write tests in an unfamiliar language
        \item To explore a new language, framework, or API
    \end{itemize}

\subsection{Experience and Feedback}

Thinking of your experience using GitHub Copilot so far, please indicate your level of agreement with the following statements:
\begin{itemize}
    \item I complete tasks faster when using Copilot
    \item I am more productive when using Copilot
    \item I spend less time searching for information or examples when using Copilot
    \item I complete repetitive programming tasks faster when using Copilot
    \item While working with an unfamiliar language, I make progress faster when using Copilot
    \item Using Copilot helps me stay in the flow
    \item Using Copilot is distracting
    \item I feel more fulfilled with my job when using Copilot
    \item I can focus on more satisfying work when using Copilot
    \item I spend less mental effort on repetitive programming tasks when using Copilot
    \item Copilot is useful
    \item Copilot is easy to use
    \item Copilot is appealing to use
    \item I learn from the suggestions Copilot shows me
    \item Getting help from Copilot while coding is easy
    \item I spend a lot of effort to understand the suggestions Copilot shows me
    \item While working with a familiar language, I make progress more slowly when using Copilot
    \item I find Copilot helpful when writing comments
    \item I find Copilot helpful when planning out code to implement
    \item I understand how to prompt Copilot so that it gives me useful responses
    \item I am concerned about the quality of my code when using Copilot
    \item I am concerned about the licensing of my code when using Copilot
    \item I find myself less frustrated during coding sessions when using Copilot
    \item The code I write using Copilot is better than the code I would have written without Copilot
\end{itemize}

6. How disappointed would you be if you could no longer use Copilot?
    \begin{enumerate}
        \item Extremely disappointed
        \item Very disappointed
        \item Somewhat disappointed
        \item A little disappointed
        \item Not disappointed at all
    \end{enumerate}

7. How likely are you to recommend that your team continues to use Copilot?
    \begin{enumerate}
        \item Extremely likely
        \item Very likely
        \item Somewhat likely
        \item Not at all likely
    \end{enumerate}

8. Is there anything else you would like to add about your experience with Copilot?
    \begin{itemize}
        \item [ ] Textbox (Please specify)
    \end{itemize}

\clearpage
\section{Appendix D: Survey Questions \& Responses based on the SPACE Framework}
\begin{figure}[h]
    \centering
    \includegraphics[width=0.9\linewidth]{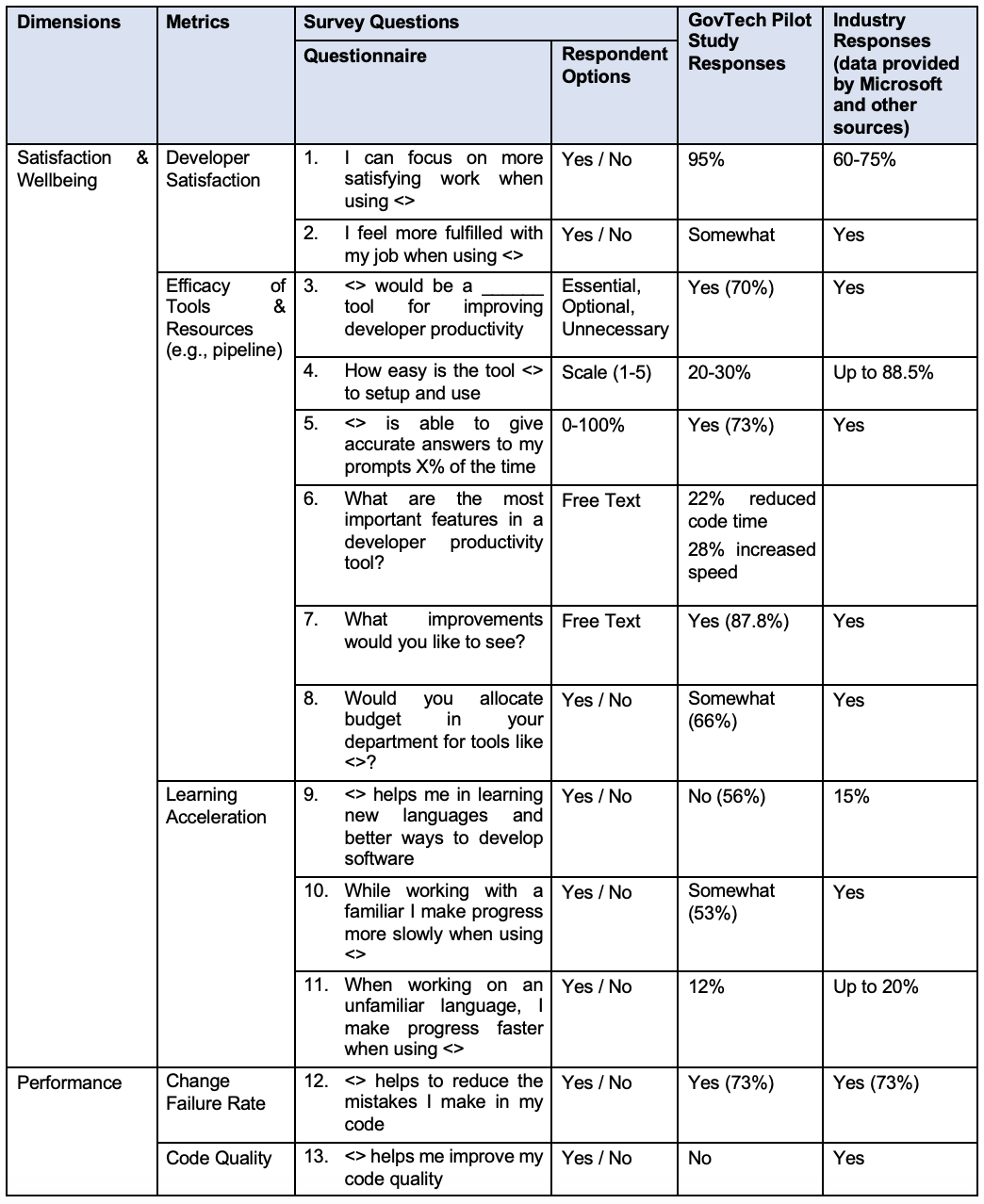}
    \label{fig:enter-label}
\end{figure}

\begin{figure}
    \centering
    \includegraphics[width=0.9\linewidth]{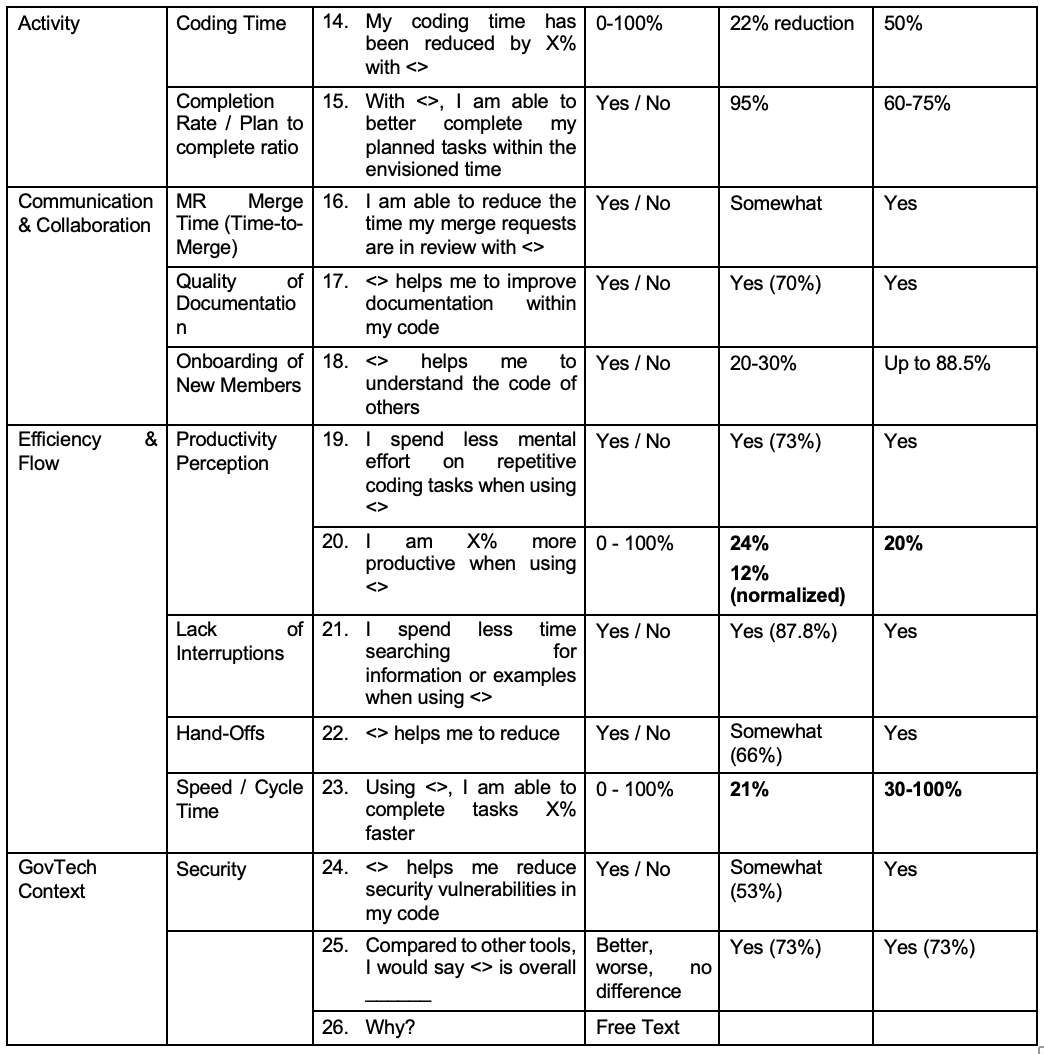}
    \caption{Survey Questions}
    \label{fig:enter-label}
\end{figure}

\begin{thebibliography}{99}

\bibitem{Gold2023}
Gold, A., Smith, J., \& Davis, M. (2023). Evaluating AI Coding Assistants in Software Development: An Industry Perspective. \textit{Proceedings of the 2023 ACM Conference on Software Engineering and Artificial Intelligence}, 45-53. \url{https://doi.org/10.1145/3597503.3608128}

\bibitem{Lee2021}
Lee, S., Wang, H., \& Davis, J. (2021). Understanding AI Assistant Use in Coding: A Review of Current Tools and Their Limitations. \textit{AI and Software Development Review}, 23(2), 122-136.

\bibitem{Mansour2022}
Mansour, H., Wu, P., \& Nguyen, K. (2022). Evaluating the Effectiveness of AI Assistants in Software Development. \textit{Journal of AI in Software Engineering}, 15(4), 205-218.

\bibitem{Imai2022}
Imai, S. (2022). "Is GitHub Copilot a substitute for human pair-programming? An empirical study," in \textit{Proceedings of the ACM/IEEE 44th International Conference on Software Engineering: Companion Proceedings}, pp. 1-4. \url{https://ieeexplore.ieee.org/document/9793778}

\bibitem{Schmidt2024}
Schmidt, T., Sergeyuk, A., Golubev, Y., \& Bryksin, T. (2024). Using AI-Based Coding Assistants in Practice: State of Affairs, Perceptions, and Ways Forward. \textit{ResearchGate}. \url{https://www.researchgate.net/publication/381372621_Using_AI-Based_Coding_Assistants_in_Practice_State_of_Affairs_Perceptions_and_Ways_Forward}

\bibitem{Pinto2023}
Pinto, G., de Souza, C., Rocha, T., Steinmacher, I., de Souza, A., \& Monteiro, E. (2023). Developer Experiences with a Contextualized AI Coding Assistant: Usability, Expectations, and Outcomes. \textit{arXiv:2311.18452}. \url{https://doi.org/10.48550/arXiv.2311.18452}

\bibitem{Wu2023}
Wu, P., Mansour, H., \& Nguyen, K. (2023). Contextual AI Assistants for Software Development: A Case Study on Internal API Integration. \textit{arXiv}. \url{https://arxiv.org/html/2404.12000v2#S4}

\bibitem{Liang2024}
Liang, J. T., Yang, C., \& Myers, B. A. (2024). A Large-Scale Survey on the Usability of AI Programming Assistants: Successes and Challenges. \url{https://doi.org/10.1145/3597503.3608128}

\bibitem{Perry2023}
N. Perry, M. Srivastava, D. Kumar, and D. Boneh, “Do users write more insecure code with AI assistants?”, 2023, \textit{arXiv:2211.03622}, \url{https://arxiv.org/abs/2211.03622}

\bibitem{Sergeyuk2024}
Sergeyuk, A., Golubev, Y., Bryksin, T., \& Ahmed, I. (2024). Using AI-based coding assistants in practice: State of affairs, perceptions, and ways forward. \textit{arXiv}. \url{https://doi.org/10.48550/arXiv.2406.07765}

\bibitem{Tan2024}
Tan, X., Long, X., Ni, X., Zhu, Y., Jiang, J., \& Zhang, L. (2024). How far are AI-powered programming assistants from meeting developers’ needs? \textit{arXiv}. \url{https://doi.org/10.48550/arXiv.2404.12000}

\bibitem{Elgedawy2024}
Elgedawy, R., Sadik, J., Dutta, S., Gautam, A., Georgiou, K., Gholamrezae, F., Ji, F., Lim, K., Liu, Q., \& Ruoti, S. (2024). Occasionally secure: A comparative analysis of code generation assistants. \textit{arXiv}. \url{https://doi.org/10.48550/arXiv.2402.00689}


\bibitem{McKinsey2023}
Deniz, B. K., Gnanasambandam, C., Harrysson, M., Hussin, A., \& Srivastava, S. (2023). Unleashing Developer Productivity with Generative AI. McKinsey Digital. \url{https://www.mckinsey.com/capabilities/mckinsey-digital/our-insights/unleashing-developer-productivity-with-generative-AI}

\bibitem{GitHubBlog1}
Kalliamvakou, E., \& GitHub Staff. (2024). Yes, good DevEx increases productivity. Here is the data. \textit{The GitHub Blog}. \url{https://github.blog/2024/01/23/yes-good-devex-increases-productivity/}

\bibitem{GitHubBlog2}
Kalliamvakou, E., \& GitHub Staff. (2023). Research: How GitHub Copilot helps improve developer productivity. \textit{The GitHub Blog}. 

\bibitem{GitHubBlog3}
Kalliamvakou, E., \& GitHub Staff. (2023). The economic impact of the AI-powered developer lifecycle and lessons from GitHub Copilot. \textit{The GitHub Blog}.

\bibitem{Arxiv2306.15033}
Golubev, Y., Sergeyuk, A., Bryksin, T., \& Ahmed, I. (2023). Sea Change in Software Development: Economic and Productivity Analysis of the AI-Powered Developer Lifecycle. arXiv:2306.15033. \url{https://arxiv.org/abs/2306.15033}

\bibitem{GetDX}
Elini. (2023). Measuring and rolling out AI coding assistants (getdx.com) – interview with head researcher Elini. \url{https://getdx.com}

\bibitem{GitHubBlog4}
Rodriguez, M. (2023). Research Quantifying GitHub Copilot's Impact on Code Quality. \url{https://github.blog/2023-10-10-research-quantifying-github-copilots-impact-on-code-quality/}

\bibitem{DevOps2023}
DevOps. (2023). Measuring GitHub Copilot's Impact on Engineering Productivity. \url{https://devops.com/measuring-github-copilots-impact-on-engineering-productivity/}

\bibitem{YouTube34}
GitHub. (2023). GitHub Copilot: The AI Pair Programmer for Today and Tomorrow. \textit{YouTube}. \url{https://www.youtube.com/watch?v=34}

\bibitem{GetDX2023}
GetDX. (2023). Developer Productivity Metrics at Top Tech Companies. \url{https://getdx.com/uploads/developer-productivity-metrics-at-top-tech-companies.pdf}

\bibitem{AWS2023}
AWS. (2023). Life at AWS: Impactful Work Helping Developers Around the World Improve Productivity. \url{https://aws.amazon.com/careers/life-at-aws-impactful-work-helping-developers-around-the-world-improve-productivity/}

\end{thebibliography}
\end{document}